\documentclass[aps,amssymb,10pt,showpacs,showkeys,letterpaper, nofootinbib]{revtex4}
\usepackage{graphicx}
\usepackage{amsmath}
\usepackage{epsfig}
\usepackage{bm}

\newcommand{\be}{\begin{equation}}
\newcommand{\ee}{\end{equation}}
\newcommand{\beq}{\begin{eqnarray}}
\newcommand{\eeq}{\end{eqnarray}}

\begin{document}

\title{Geodesic Structure of the Schwarzschild Black Hole in Rainbow Gravity }

\author{ Carlos Leiva }
 \email{cleivas@uta.cl}
\affiliation{\it Departamento de F\'{\i}sica, Universidad de
Tarapac\'{a}, Casilla 7-D, Arica, Chile}

\author{Joel Saavedra}
 \email{ Joel.Saavedra@ucv.cl }
 \affiliation{\it Instituto de F\'{\i}sica, Pontificia Universidad
de Cat\'{o}lica de Valpara\'{\i}so, Casilla 4950, Valpara\'{\i}so, Chile}

\author{ Jos\'{e} Villanueva }
 \email{jose.villanueva.l@mail.ucv.cl}

\affiliation{\it Instituto de F\'{\i}sica, Pontificia Universidad
de Cat\'{o}lica de Valpara\'{\i}so, Casilla 4950, Valpara\'{\i}so
, Chile}

\date{\today}

\begin{abstract}

  In this paper we study  the geodesic structure of the Schwarzschild black hole in rainbow gravity analyzing the behavior of null and time-like
geodesic. We find that the structure of the geodesics essentially does not
change when the semi-classical effects are included. However, we can distinguish  different scenarios if we take
into account the effects of rainbow gravity. Depending on
the type of rainbow functions under  consideration, inertial and
external observers  see very different situations in radial
and non radial motion of a test particles.
\end{abstract}



\maketitle
\section{\label{sec:Int}Introduction}

Nowadays one of the most challenging tasks in Theoretical Physics is to
combine quantum theory and general relativity. Several lines of research
have been developed but  none of them is completely
successful in obtaining a complete description of the quantum gravity
realm. Meanwhile, some phenomenological approaches have been put
on the table. One of them is the modification of the dispersion
relation $E^2-p^2=m^2$ \cite{piran}, with a non linear version
instead. After all life is not linear at all, so it is very
probable that the linear version of the relation linking energy
and momenta is just a first approximation to a real non lineal
one. This assumption is a  base of quantum gravity models,
that suggests that it could be desirable to review the Lorentz
invariance relations and the very structure of the space time at
high energy scales, and not  wipe
Lorentz symmetry out, but  just to modify it to have a non linear version
that fits with the usual one at low energies.

On the other hand, some data seem to invite to introduce a minimal
length in physical theories. Indeed, there already exist  well established
theories, such  as String Theory or Loop Quantum Gravity, that have some
fundamental quantities: the Planck longitude $l_p=\sqrt{\hbar
G/c^3}$, the associated time scale $t_p=l_p/c$ and the Planck
energy $E_p=\hbar /t_p$. All of them suppose that beyond these
thresholds, the physics should change dramatically. However, even
discreteness is, in some models,  coherent with Lorentz symmetry, since these absolute
values of longitude, time or energy are not in total agreement
with the Lorentz transformations and  this fact is additional motivation to modify the Lorentz boosts.

In fact, among all proposals  to deepen  our  understanding of
the nature of spacetime  by changing some apparently well settled ideas
in Physics,  we can find a very interesting one that is to modify
the Lorentz boosts through the Double Special relativity (DSR)
proposals \cite{mag,giovanni2,kow}. These theories are based on a
generalization of Lorentz transformations through a more broad
point of view of conformal transformations, where there exist  two observer
independent scales, velocity of light and Planck length. These
theories are rather polemical, but are of increasing interest too
because they can be useful as  effective new tools in gravity
theories for example, in Cosmology as an alternative to inflation
\cite{mof,alb}, or in other fields like propagation of light
\cite{ku}, that are related, for instance, to cosmic microwave
background radiation.

Finally, it is worthy to study an effect of having a modified
dispersion relation (that we expect represents quantum effects),
in a strong gravitational field, such as a black hole. In this
approach, there are several works about  called rainbow gravity,
whose history begins more or less with a treatment done in
Ref.\cite{Magueijo:2002xx}. In particular, it could be interesting
in near future, study about  the effects of this quantum
corrections here proposed,  on the Schwarzchild black hole lensing
features that were well reviewed on \cite{VE} and \cite{V}, in the
standard case, in order to verify the real existence of the
deformations predicted from rainbow gravity formalism.

In this paper we review the structure of geodesics near a
Schwarzschild black hole, motivated by the fact that black holes provide gravity
conditions  to test quantum
effects due to the discrete nature of spacetime or the existence of
a limit in the energy that a particle can bear. If a testable
effect is encountered, we could have a reliable way to examine an
underlying hypothesis of a modified dispersion relation.

The paper is organized as follows: In the next section we
introduce the rainbow gravity that is a modified version of
the Schwarzschild solution of Einstein equation  due to generalized
dispersion relations above mentioned. In the third section we
analyze the geodesic structure, and in the fourth one we discuss
conclusions.

\section{\label{sec:Dilatonic}Rainbow Gravity}

Rainbow gravity was proposed in Ref.\cite{Magueijo:2002xx} and it is  based
on two principles. The first one is the correspondence principle that provides validity of
standard general relativity in the limit of low energies relative
to the Planck Energy. Then, when  $E/E_{pl}<<1$, rainbow gravity
becomes the standard general relativity. The  other one is the Modified
Equivalence Principle, that ensures to have a freely
falling observer that measures the same laws of physics  as in
modified special relativity. Therefore, we can construct an energy
dependent orthonormal frames locally given by
\begin{equation}
g(E)=\eta ^{ab}e_{a}(E)e_{b}(E),  \label{metric1}
\end{equation}
where  the correspondence principle is established through the
explicit expression for the orthonormal base
\begin{eqnarray}
e_{0}(E) &=&\frac{1}{f(E/E_{pl})}\widetilde{e}_{0},  \label{frame} \\
e_{i}(E) &=&\frac{1}{g(E/E_{pl})}\widetilde{e}_{i}.  \nonumber
\end{eqnarray}
Having Minkowski
space in the low energy limit  $E/E_{pl}\rightarrow 0$ implies the following relation between the
arbitrary functions $f(E/E_{pl})$ and  $g(E/E_{pl})$
\begin{equation}
\lim_{E/E_{pl}\rightarrow 0}f(E/E_{pl})=\lim_{E/E_{pl}\rightarrow
0}g(E/E_{pl})=1.  \label{limit}
\end{equation}
On the other hand, arbitrary energy dependent function
originates in  the modification of Lorentz dispersion relation
\begin{equation}
E^{2}f\left( E/E_{pl}\right) -p^{2}h\left( E/E_{pl}\right)
=m_{0}^{2}, \label{dispersion1}
\end{equation}
that explicitly shows a deformed nature,  or doubly special relativity,  as a
class of theories that implement a changed set of principles to
special relativity.
>From the above equations  we see that the metric
(\ref{metric1}) that describes a flat rainbow space, can be
generalized to a curve rainbow space. Similarly as in the flat case, it
corresponds to a one-parameter family of metrics given in terms of  one-parameter family of orthonormal frame fields (\ref{frame}), whose
energy-dependent metric must satisfy a modified version of Einstein
Equations given by
\begin{equation}
G_{\mu \nu }(E)=8\pi G_{N}(E)T_{\mu \nu }(E)+g_{\mu \nu
}(E)\Lambda (E). \label{einsteineq}
\end{equation}
Here, the Newton constant $G_{N}(E)$ and the cosmological constant
(for asymptotically Anti de Sitter or de Sitter space) $\Lambda
(E)$ are allowed to be energy-dependent, and they satisfy  the
correspondence principle.

In this approach  in Ref. \cite{Magueijo:2002xx}, it was presented a
modified general spherically symmetric solution to equations
(\ref{einsteineq}),  known as Schwarzschild modified Black
Hole, described by the metric
\begin{equation}
ds_{Schw}^{2}=-\frac{(1-\frac{2G(0)M}{\widetilde{r}})}{f^{2}(E/E_{pl})}d%
\widetilde{t}^{2}+\frac{1}{(1-\frac{2G(0)M}{\widetilde{r}})g^{2}(E/E_{pl})}d\widetilde{r}^{2}+\frac{\widetilde{r}^{2}}{g^{2}(E/E_{pl})}d\widetilde{%
\Omega }^{2},  \label{Schw}
\end{equation}
where the quantities $(\widetilde{t}, \widetilde{r},
\widetilde{\Omega})$ are independent energy variables. The
metric  also depends on the energy of a particle moving in it. That
is, two different test particles observe different effective space-time geometries. As a consequence,  the present space-time is
endowed with a Plank-scale modification that carries some quantum
effects. We are  interested in  studying  a behavior of a test
particle under the influence of this geometry with quantum
corrections. This point will be analyzed  in next section.

\section{Geodesics}

Using the variational principle \cite{{Chandrasekhar},{COV}}, the metric (\ref{Schw}) is associated with  a Lagrangian density $\mathcal{L}$ given by
\be 2
\mathcal{L}=-\frac{F(r)}{f_{E}^{2}}\dot{t}^{2}+\frac{1}{g_{E}^{2}
F(r)}\dot{r}^{2}+\frac{r^{2}}{g_{E}^{2}}\dot{\Omega^{2}},
\label{1} \ee
where $f_{E} \equiv f(E)$ and  $g_{E} \equiv g(E)$,
 whereas $F(r)$ is the usual lapses  function of the Schwarzschild spacetime
\footnote{This treatment is valid for the Kottler spacetime if
cosmological constant does not depend on the constant of motion
$E$, in which case a complete analytic solution of the geodesic
structure of the Schwarzschild anti-de Sitter was done in
\cite{COV}, with the lapses function given by
$F(r)=1-\frac{r_{+}}{r}+\frac{r^{2}}{\ell^{2}}$, where
$\frac{\Lambda}{3}=-\frac{1}{\ell^{2}}$.} \be
F(r)=1-\frac{r_{+}}{r},\label{2} \ee and
$\dot{\Omega^{2}}=\dot{\theta}^{2}+sin^{2}\theta \dot\phi$. In
this notation  dot represents a derivative with respect to proper
time, $\tau$ (affine parameter along a geodesic). Since the
Lagrangian does not depend on ($t, \phi$), the corresponding
conjugate momenta are conserved, therefore  in the invariant plane
$\theta = \pi/2$ we have \be \Pi_{t} = -\frac{F(r)}{f_{E}^{2}}
\dot{t} = - E, \label{3}\ee and \be \Pi_{\phi} =
\frac{r^{2}}{g_{E}^{2}} \dot{\phi} = L, \label{4}\ee where $E$ and
$L$ are constants of motion. From the last two equations, and
taking $\bar{\tau} = g_{E} \tau$, $\mathbb{E} = f_{E} E$ and
$\mathbb{L} = g_{E} L$, the Lagrangian (\ref{1}) can be written in
the following form \be 2\mathcal{L}\equiv
-m=-\frac{\mathbb{E}^{2}}{F(r)}+\frac{\dot {\bar{r}}^{2} }{
F(r)}+\frac{\mathbb{L}^{2}}{r^{2}}, \label{5} \ee where, by
normalization,  $m=1$ for massive particles (time-like geodesics)
and $m=0$ for massless particles (null geodesics), and
$\dot{\bar{r}}=dr/d\bar{\tau}$. Thus, our equation of motion
becomes \be \dot{\bar{r}}^{2}=
\mathbb{E}^{2}-\mathbb{V}_{G}(r;m,\mathbb{L}), \label{6}\ee where
$\mathbb{V}_{G}(r;m,\mathbb{L})$ is the generalized effective
potential, which is given by \be
\mathbb{V}_{G}(r;0,\mathbb{L})\equiv \mathbb{V}_{N}(r;\mathbb{L})
= F(r) \frac{\mathbb{L}^{2}}{r^{2}},\label{7}\ee for null
geodesics, and \be \mathbb{V}_{G}(r;1,\mathbb{L})\equiv
\mathbb{V}_{T}(r;\mathbb{L}) = F(r)
\left(1+\frac{\mathbb{L}^{2}}{r^{2}}\right),\label{8}\ee for
time-like geodesics. In what follows, we use obtained results to
discuss two families of functions, say, DSR1 with $f_{E}=1$ and
$g_{E}=1+\frac{1}{2} l_{p} E$; and DSR2 with $f_{E}=g_{E}
(=1+\frac{1}{2} l_{p} E)$ \cite{mendez}.

\subsection{Null Geodesics}

 Related to the equation of motion for massless particle, we start from Eqs.  (\ref{6}) and  (\ref{7}),  and we study independently  the radial and non-radial motion.

\medskip

(a.i).- \underline{\underline{Radial Null Geodesics}}

\medskip

\noindent In this case we have $\mathbb{V}_{N}(r;\mathbb{L}) =0$ and the radial motion is
governed by
\be  \dot{\bar{r}}^{2}=\mathbb{E}^{2}, \label{13}\ee
therefore
\be \Delta \tau=\frac{(\Delta r/E)}{\Gamma_{1}}=\frac{\Delta
\tau_{Schw}}{\Gamma_{1}},\label{14}\ee
where $\Gamma_{1}= g_{E} f_{E}$. We see that the radial motion
of  massless particle shows the same behavior as standard Schwarzschild geometry,  with the only difference that its
proper time is rescaled  by a factor $\Gamma_{1}$.

\begin{figure}[!h]
  \begin{center}
    \includegraphics[width=100mm]{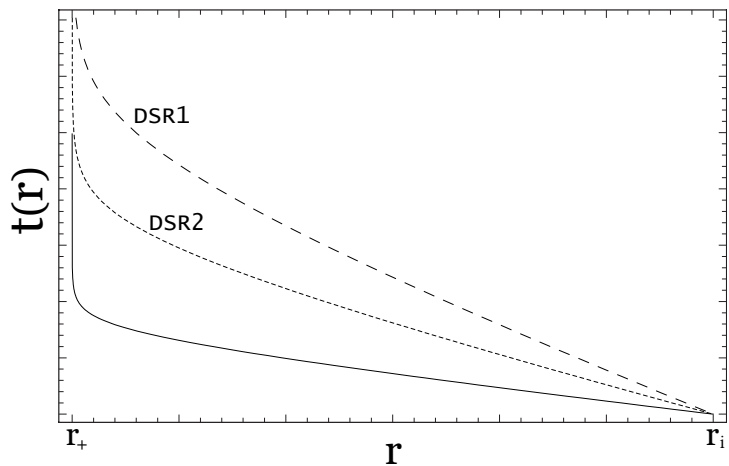}
  \end{center}
  \caption{This plot shows the coordinate time $t$ versus the radial distance from the black hole, $r$, for massless particles falling in radial motion. The solid curve represents  the semi-classical limit (Schwarzschild case). The dashed curves represents  the modifications DSR1 and DSR2 in the limit $E \rightarrow E_{p} (=l_{p}^{-1})$.}
  \label{fig:trng}
\end{figure}

Furthermore, from Eqs. (\ref{3}) and (\ref{13}), we find an expression for the
coordinate time, $t$,
\be \frac{dt}{dr}=\frac{\Gamma_{2}}{F(r)},
\label{15}\ee
in which case we obtain
\be \Delta t= \Gamma_{2} \Delta t_{Schw} = \Gamma_{2} \left[(r_{i}-r) +
r_{+} \log \left(\frac{\frac{r}{r_{+}} - 1}{\frac{r_{i}}{r_{+}} - 1} \right) \right],\label{16}\ee
where $\Gamma_{2}=\frac{g_{E}^{3}}{f_{E}}$. This
situation is showed in Fig. \ref{fig:trng} in case of DSR1 and DSR2, in the limit $E \rightarrow E_{p} (=l_{p}^{-1})$, together with the semi-classical limit (Schwarzschild case).

\medskip

(a.ii).- \underline{\underline{Non-Radial Null Geodesics}}

\medskip

Returning to the equation of motion (\ref{6}-\ref{7}), it is convenient to
rewrite it in terms of the new variable $u=1/r$, and using (\ref{4})
it can be put in the form analog to the one shown in Ref.\cite{Chandrasekhar}
\be \left( -\frac{du}{d \phi}\right)^{2}=r_{+} u^{3} - u^{2} +
\frac{1}{\bar{b}^{2}}.\label{17}\ee
Here  $\bar{b} = \sqrt{\Gamma_{3}} b$ denotes the generalized impact
parameter for the orbital motion of the massless particles, and $\Gamma_{3}=\Gamma_{2} / \Gamma_{1} = g_{E}^{2}/
f_{E}^{2}$. Note that, from the equation (\ref{17}), the non-radial motion is
identical to the one corresponding to the Schwarzschild black hole when the DSR2 deformation is considered.

\subsection{Time-Like Geodesics}

The motion of a  massive particle is described by Eqs.  (\ref{6}-\ref{8})
\be  \dot{\bar{r}}^{2} =\mathbb{E}^{2} - F(r) \left(1 +
\frac{\mathbb{L}^{2}}{r^{2}}\right), \label{18}\ee

(b.i).- \underline{\underline{Radial Time-Like Geodesics}}

\medskip

In this case, the equation of motion (\ref{18}) can be written as
\be  \left(\frac{dr}{d\bar{\tau}}\right)^{2}
=\frac{r_{+}}{r}-(1-\mathbb{E}^{2}). \label{rtl1}\ee Making the
usual substitution \be R = \frac{r}{r_{i}} =
\cos^{2}\frac{\eta}{2} ,\label{rtl2}\ee where ($R_{+} =
\frac{r_{+}}{r_{i}} < u \leq 1$), we obtain for the proper time
and coordinate time respectively \be \tau=\frac{r_{i}}{g_{E}}
\Theta_{l1} ,\label{rtl3}\ee \be  t = r_{i}
\frac{f_{E}\mathbb{E}}{g_{E}}\Theta_{l2} ,\label{rtl4}\ee where we
introduce the functions \be \Theta_{l1} =
\frac{1}{\sqrt{R_{+}}}\left[\arccos \sqrt{R} +  \sqrt{R - R^{2}}
\right], \label{rtl5}\ee and \be \Theta_{l2} =
\frac{1}{\sqrt{R_{+}}}\left[ \sqrt{R - R^{2}} + (2 R_{+} +
1)\arccos\sqrt{R} + \frac{2 R_{+}^{2}}{\sqrt{R_{+} - R_{+}^{2}}}
\text{arctanh}\sqrt{\frac{R_{+}(1 - R)}{R(1-R_{+})}}  \right].
\label{rtl6}\ee The returning point  $r_{i} = \frac{r_{+}}{1 -
\mathbb{E}^{2}}$  was chosen as the starting point, $r_{0}=r_{i}$.
To have positive $r_{i}$, we have to impose $\mathbb{E}^{2}<1$,
and  we shall assume that $E \lesssim 1$ (Schwarzschild case $E <
1$). This means that energies are low $E \lll E_{p}$ and the
corrections from the rainbow gravity are negligible.

Consider now a sector out of the capture zone, i.e, where $\mathbb{E}^{2}>1$. In this case, the change of radial variable suggested by (\ref{rtl1}) is
\be R = \frac{r}{r_{e}} = \sinh^{2}\frac{\xi}{2} , \label{rtl7}\ee
 where $r_{e} = \frac{r_{+}}{\mathbb{E}^{2} - 1}$ is a distance-energy parameter of a falling particle. Therefore, for the  particle falling from the distance  $r_{0}$ ($R_{0}=r_{0} / r_{e}$), we have
\be \tau=\frac{r_{e}}{g_{E}}  \left[\Theta_{g1}(R) - \Theta_{g1}(R_{0}) \right],\label{rtl8}\ee
and
\be t= r_{e} \frac{f_{E}\mathbb{E} }{g_{E}} \left[\Theta_{g2}(R) - \Theta_{g2}(R_{0}) \right],\label{rtl9}\ee
where
\be \Theta_{g1}(R)=\frac{1}{\sqrt{R_{+}}}\left[ \text{arcsinh}\sqrt{R}  -
\sqrt{R + R^{2}}  \right],\label{rtl10}\ee
and
\be \Theta_{g2}(R)=\frac{1}{\sqrt{R_{+}}}\left[(2R_{+} + 1)\text{arcsinh}\sqrt{R} - \sqrt{R + R^{2}}  + \frac{4 R_{+}^{2}}{\sqrt{R_{+} + R_{+}^{2}}} \text{arctanh}\sqrt{\frac{R_{+}(1 - R)}{R(1-R_{+})}} \right].\label{rtl11}\ee

\begin{figure}[!h]
  \begin{center}
    \includegraphics[width=140mm]{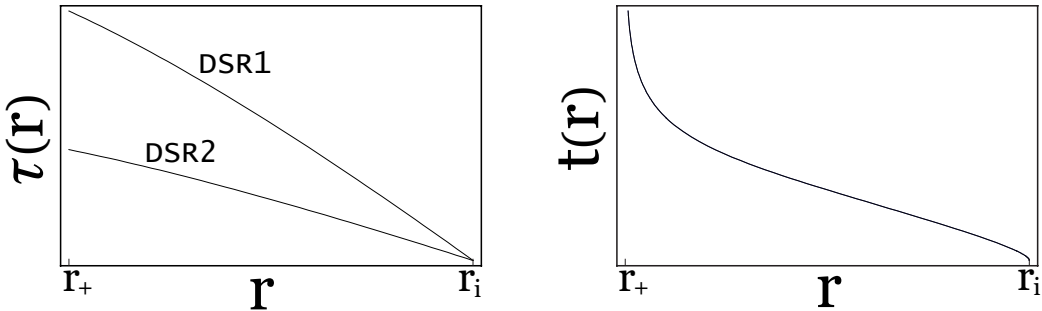}
  \end{center}
  \caption{This plot shows a  behavior  of a  massive particle  radially  falling into the black hole in the region out of the capture zone, i.e. when $E>1$. Left panel shows the proper time $\tau$ as a function of a distance $r$, and it has larger value in the  DSR1 than DSR2  scenario in the limit $E \rightarrow E_{p} (=l_{p}^{-1})$. Right panel shows coordinate time $t$ versus radial coordinate $r$. In the limit $E \rightarrow E_{p} $, both cases of DSR1 and DSR2 scenarios, converge to the same function. This means that an outside observer can not see differences between the DSR1 and DSR2 modifications.}
  \label{fig:ttlg}
\end{figure}

In Fig. \ref{fig:ttlg} we show the proper and coordinate times for radial massive particle freely falling  into the black hole. Remarkable feature is that, in comparison with the null radial motion, the DSR1 and DSR2 modifications for time-like motion coordinate times, both merged in one curve.

(b.ii).- \underline{\underline{Non-Radial Time-Like Geodesics}}

Using (\ref{8}) and defining  $r=1/u$, the motion of a non-radial
massive particle is governed by equation \be
\left(\frac{du}{d\phi}\right)^{2} =r_{+} u^{3} - u^{2} +
\frac{r_{+}}{\mathbb{L}^{2}} u - \frac{1 -
\mathbb{E}^{2}}{\mathbb{L}^{2}}. \label{nrtl1}\ee Again, if we
consider lower energies ($\mathbb{E} \sim E$ and $\mathbb{L} \sim
L$), we obtain the  corrections from rainbow gravity   are
negligible. This means that  bounded orbits for massive particles
are not affected by the effects of  rainbow gravity.

\section{\label{sec:Dilatonic2}Discussion and Outlook}

We study the geodesic structure of Schwarzschild black holes in
rainbow gravity, analyzing the behavior of null and time-like
geodesics for DSR1 and DSR2 theories. We found that the structure
of geodesics does not change when  semi-classical effects are
taken into account. The case of radial null-geodesics shows that
the effects of a space-time endowed with a Planck-scale
modification, and therefore including the quantum effects, are of
kinematic origin and the only correction is an  adding  an another
proper time contraction equation (\ref{14}) . For the coordinate
time Eq.(\ref{16}), in both theories we found larger values of
times than in  standard Schwarzschild case. Our results in that
case are summarized in Fig. 1,  for both DSR1 and DSR2. The
photons with energy of order of the Planck scale exhibit a
modification in the Doppler effect, as seen from outside. For a
non radial-null geodesic, the only modification comes from the
impact parameter $\overline{b}=\sqrt{g_E^2/f_E^2}\,b_{schw}$. In
the case of a massive particle we found, for radial geodesics,
that the only modification is in a  changed  returning point,
whereas for a non-radial geodesic, the effects depend on the
relation between $g(E)$ and $f(E)$ under consideration.

Therefore, we conclude that different test particles (with
different energies) do not see different spacetimes. Based on our
results, different test particles
have  different effective descriptions, that means that there are some changes
in their kinematics properties around the same
rainbow Schwarzschild black hole when the quantum effects have
been taken in to  account.

\begin{acknowledgments}

We are grateful to O. Mi\v{s}kovi\'{c} for careful reading of the
manuscript. The authors acknowledge the referee for useful
suggestions in order to improve the presentation of the results of
this paper. J. S. was supported by COMISION NACIONAL DE CIENCIAS Y
TECNOLOGIA through FONDECYT Grant 11060515 and by PUCV Grant N$^0$
123.789. J. V. was founded by MECESUP FSM 0204 . C.L. was
supported by Grant UTA DIPOG N$^0$ 4721-07.
\end{acknowledgments}

\end{document}